\documentclass[aps,prl,twocolumn,groupedaddress,floats,showpacs]{revtex4}
\usepackage{latexsym}
\usepackage{dcolumn}
\usepackage[dvips]{graphicx}
\usepackage{amssymb}
\usepackage{graphics}
\usepackage{amsmath}
\usepackage{epsf}

\newcommand{\vk}{{\bf k}}

\begin{document}

\title{Exotic plasmon modes of double layer graphene}
\author{E. H. Hwang and S. \ Das Sarma}
\address{Condensed Matter Theory Center, Department of physics,
  University of Maryland, College Park, 
Maryland  20742-4111}
\date{\today}

\begin{abstract}
We derive the plasmon dispersion in doped double layer graphene (DLG),
made of two parallel graphene mono-layers with carrier densities
$n_1$ and $n_2$ respectively, and an inter-layer separation of
$d$. The linear chiral gapless single particle energy dispersion of
graphene leads to exotic DLG plasmon properties with several
unexpected experimentally observable characteristic features such as a
nontrivial influence of an undoped ($n_2=0$) layer on the DLG plasmon
dispersion and a strange influence of the second layer even in the
weak-coupling $d\rightarrow 0$ limit. Our predicted exotic DLG plasmon
properties clearly distinguish graphene from the extensively studied
usual parabolic 2D electron systems.
\end{abstract}


\maketitle

Graphene, a two dimensional (2D) layer of carbon atoms arranged in a
honeycomb lattice, has attracted a great deal of attention because of
its unique electronic properties arising from its chiral, gapless,
linear band dispersion \cite{review}. Of particular importance and
fundamental 
interest are those graphene properties which are {\it qualitatively}
different from the behavior of electrons (or holes) in the extensively
studied 2D semiconductor systems. Some of these unique graphene
properties \cite{review} are its 'half-integral' integer quantized
Hall effect, a 
carrier mobility independent of carrier density away from the charge
neutrality point (i.e. the Dirac point), and the finite minimum
conductivity around the Dirac point without any obvious
metal-insulator transition. In this Letter, we predict a new exotic
collective property of graphene electrons, which distinguishes
graphene qualitatively from the semiconductor-based parabolic 2D
electron systems. We find the surprising result that the plasma modes
of an interacting double layer graphene (DLG) system are completely
different from the well studied double layer semiconductor quantum
well plasmons \cite{dassarma}. Our finding is unexpected and
interesting because the 
collective plasmon excitations arise from the long-range Coulomb
interaction which is the same in graphene and
parabolic 2D systems. The exotic DLG palsmon properties are
experimentally observable and
should have important consequences for the many-body effects in
doped multi-layer graphene structures.

Plasma modes of 2D double layer structures have been extensively
studied ever since the
existence of an undamped acoustic plasmon mode was predicted
\cite{dassarma} in semiconductor double quantum well systems.
It is known that 
when the two 2D layers are put in close proximity with 
a high insulating barrier between them to prevent interlayer
tunneling, the 2D plasmons are coupled by
the interlayer Coulomb interaction leading to the formation
\cite{dassarma} of two
branches of longitudinal collective excitation spectra called
the optical plasmon (OP, $\omega_+ \sim q^{1/2}$, where $q$ is the 2D
wave vector) and the acoustic
plasmon (AP,  $\omega_-\sim q$), 
where the density fluctuations in each component oscillate in
phase (OP) and out of phase (AP), respectively, relative to
each other, assuming the two charge components to be the same.
These collective modes of the double layer structures, which have been
directly experimentally observed \cite{r1}, play important
roles in the many body properties such as screening
and  drag  \cite{drag,drag_eh}.

Here we consider the
experimentally relevant issue of the collective mode dispersion in
DLG formed by two parallel single layer graphene (SLG) separated by a
distance $d$.
The DLG is fundamentally different from well studied
bilayer graphene \cite{BLG} because there is no inter-layer tunneling,
only inter-layer Coulomb interaction.
Spatially separated two-component DLG can be
fabricated by folding an SLG over a high insulating
substrate \cite{r2}.

We calculate the DLG plasmon modes and their loss functions (spectral
strength).
We find many intriguing and unexpected features of coupled plasmon
modes and their Landau damping in DLG.
Even though we recover in the long wavelength ($q \rightarrow 0$)
limit the well-known optical 
($\omega_+ \sim q^{1/2}$) and acoustic ($\omega_- \sim q$) plasmons,
the density dependence is given by 
$\omega_{+}^2(q \rightarrow 0)  \propto  \sqrt{n_1}+\sqrt{n_2}$ and 
$\omega_-^2(q \rightarrow 0)  \propto  { \sqrt{n_1
    n_2}}/({\sqrt{n_1} +\sqrt{n_2}})$
compared to  $\omega_+^2 \propto N$ and
$\omega_-^2 \propto n_1 n_2/N$ in the ordinary 2D system,
where $N=n_1+n_2$.
When the interlayer Coulomb coupling is weak
(i.e. for large separations, $k_Fd\gg 1$) the undamped DLG modes
($\omega_{\pm}$) become
degenerate and have the same frequency as the 
uncoupled SLG plasmon \cite{hwang}, but inside the
interband Landau damping regime this degenerate mode splits into
two coupled modes. In the strong coupling regime
($k_Fd \ll 1$) the $\omega_-$ mode
approaches the intraband electron-hole continuum (i.e. $\omega_-
\rightarrow v_F q$) and is overdamped,
but the $\omega_+$ mode is shifted to higher frequency and in the long
wavelength limit  becomes the SLG plasmon with a density 
$N= n_1 + n_2 +2\sqrt{n_1 n_2}$ 
instead of the  density $N=n_1+n_2$ as 
happens in regular 2D double layer systems.
We also find that the DLG plasmon modes are heavily damped even in the
long wave length limit when the two layers have unequal density,
even when one layer is undoped (i.e. $n_2=0$).
We believe that our predictions
may be easily observable via inelastic light
scattering spectroscopy \cite{pinczuk}, frequency-domain far
infrared (or microwave) spectroscopy \cite{allen}, or inelastic
electron scattering spectroscopy \cite{R1,Liu}.

The plasmon modes 
are given by the poles of the density-density correlation
function, or equivalently by the zeros of the dynamical dielectric
function.  For a double layer system the collective
modes are given by the zeros of the generalized dielectric
tensor $\epsilon_{lm}$ where {\it l,
m}$=$1 or 2 with 1,2 denoting the two
layers. \cite{dassarma}  The
dielectric function $\epsilon_{lm}(q,\omega)=\delta_{lm} - v_{lm}
\Pi_m$ is obtained 
within the mean field random phase approximation (RPA) in our theory,
and the two component determinantal equation becomes
\begin{eqnarray}
\epsilon(q,\omega)& = & [1-v_{11}(q)\Pi_{1}(q,\omega)]
	[1-v_{22}(q)\Pi_{2}(q,\omega)] \nonumber \\
        &-&
	v_{12}(q)v_{21}(q)\Pi_{1}(q,\omega) \Pi_{2}(q,\omega).
\label{em}
\end{eqnarray}
where
$v_{ll}(q)$ and $v_{lm}(q)$ are respectively the intralayer and
interlayer Coulomb interaction matrix elements. In our model,
$v_{ll}(q) = 2\pi e^2/(\kappa q)$; $v_{lm}(q) = v_{ll}(q)
\exp(-qd)$, with $\kappa$ as the background lattice
dielectric constant. 
Finally, $\Pi_{l}(q,\omega)$ in
Eq. (\ref{em}) is the
noninteracting SLG polarizability function
given by  \cite{hwang}: 
\begin{equation}
\Pi_l(q,\omega) =-\frac{g}{L^2}
\sum_{{\bf k}ss'}\frac{f_{s{\bf k}}-f_{s'{\bf k}'}}
{\omega + \epsilon_{s{\bf k}}-\epsilon_{s'{\bf k}'} +i\eta}F_{ss'}({\bf
  k},{\bf k}'),
\label{pol}
\end{equation}
where $g=g_sg_v$ (with $g_s=2$ and $g_v=2$ being spin and valley
  degeneracy),  ${\bf k}'={\bf k}+{\bf 
  q}$, $s,s'=\pm 1$ denote the band indices, $\epsilon_{s\vk}=s v_F
  |{\bf k}|$ ($v_F$ being the Fermi velocity of graphene and $\hbar=1$
  throughout this paper),
and $F_{ss'}({\bf k},{\bf k}')$ is
the overlap of states and given by
$F_{ss'}({\bf k},{\bf k}') = \frac{1}{2}(1 + ss' \cos\theta),$
where $\theta$ is the
angle between ${\bf k}$ and ${\bf k}'$, and
$f_{sk}$ is the Fermi distribution function, 
$f_{s{\bf k}} = [\exp \{\beta(\epsilon_{s{\bf k}}-\mu_l)\} + 1]^{-1}$,
with $\beta = 1/k_BT$ and  $\mu_l$ the chemical potential of $l$-the
layer. 
In the long wavelength limit ($q\rightarrow 0$) the polarizability
  becomes \cite{hwang}
\begin{equation}
\Pi_l(q,\omega)=\frac{g k_{F_l}}{4\pi v_F} \frac{q^2}{\omega^2} \left [
  1-\frac{\omega}{4E_{F_l}}\ln \frac{2E_{F_l}+\omega}{2E_{F_l}-\omega}
  \right ]. 
\label{pol_l}
\end{equation}
where $k_{F_l}$ ($E_{F_l}=v_F k_{F_l}$) is the Fermi wave vector (Fermi energy)
of the $l$-th layer.

The analytical formula for the long-wavelength plasmon
dispersion can be obtained in the general situation
where the two layer have different Fermi wave
vectors and Fermi energies by virtue of having
different  densities $n_1,n_2$. 
From Eqs.~(\ref{em}) and (\ref{pol_l}) 
it is possible to obtain the following long
wavelength  plasma modes of the coupled DLG:
\begin{subequations}
\begin{eqnarray}
\omega_{+}^2(q \rightarrow 0) & = & 2r_s v_F^2(k_{F_1}+k_{F_2}) q, \\
\omega_-^2(q \rightarrow 0) & = & \frac{4r_s v_F^2 k_{F_1}k_{F_2}d}
      {k_{F_1}+k_{F_2}} q^2,
\end{eqnarray}
\end{subequations}
where $r_s =
e^2/\kappa v_F$ is the effective graphene fine structure constant. 
Thus, one recovers in a straightforward fashion the well-known optical
($\omega_+ \sim q^{1/2}$) and acoustic ($\omega_- \sim q$) plasmons of
a double layer system.
In addition we have the density dependence of 
DLG plasmons as 
$\omega_{+}^2(q \rightarrow 0)  \propto  \sqrt{n_1}+\sqrt{n_2}$ and 
$\omega_-^2(q \rightarrow 0)  \propto  { \sqrt{n_1
    n_2}}/({\sqrt{n_1} +\sqrt{n_2}})$, which are different from the
density dependence of the corresponding plasmon dispersion 
in ordinary 2D systems \cite{dassarma} where the $\omega_+^2$ 
depends on the total 2D electron density $N=n_1+n_2$,
(i.e. $\omega_+^2 \propto N$), and 
$\omega_-^2 \propto n_1 n_2/N$.

\begin{figure}
\bigskip
\epsfxsize=.95\hsize
\epsffile{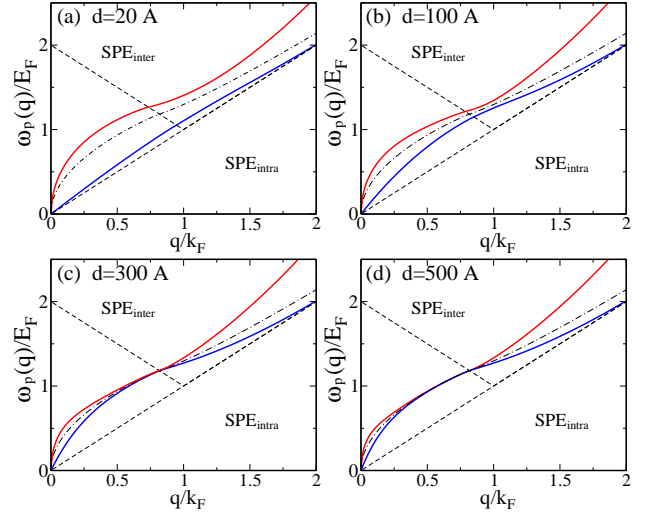}
\caption{\label{fig0}
Calculated plasmon mode dispersions of coupled DLG
for a fixed  density and several layer separations.
The upper (lower) solid line indicates the in-phase mode $\omega_+$
(the out of phase mode $\omega_-$).
Here we use the parameters: $n_1 = n_2 = 10^{12}cm^{-2}$ and (a)
$d=20$ \AA \; ($k_F d = 
0.35$), (b) $d=100$ \AA \; ($k_F d = 1.8$), (c) $d=300$ \AA \; ($k_F d =
5.3$), (d) $d=500$ \AA \; ($k_F d = 8.9$). The dot-dashed line indicates the
plasmon mode dispersion of SLG with the same
density. The dashed lines indicate the boundaries of SPEs (Landau
damping region for intra- and interband electron-hole excitations).
}
\end{figure}

In Fig.~\ref{fig0} we show the calculated DLG plasmon dispersions
($\omega_{\pm}$) for equal densities of $n_1 = n_2 =
10^{12}cm^{-2}$ and several layer separations.
The plasmon mode dispersion of the uncoupled SLG
with the same density is shown as the dot-dashed line.
In Fig.~\ref{fig0}, we also
show the electron-hole continua or single-particle excitation
(SPE) region, which determines the decay
(Landau damping) of the plasmon at a given frequency
and wave vector. 
If the collective mode lies inside the
SPE continua, we expect the mode to be Landau damped.  
The SPE continuum is defined by
the nonzero value of the imaginary part of the total dielectric
function, Im$[\epsilon(q,\omega)] \neq 0$, because the damping of
the plasmon by emitting an electron-hole single pair excitation is
allowed in the region of nonzero Im[$\Pi(q,\omega)$].
For graphene, both intraband and interband SPE transitions are
possible \cite{hwang}, and the boundaries (dashed lines) are given in
Fig.~\ref{fig0}.  
Due to the phase-space restriction, the interband SPE continuum opens a
gap in the long wavelength region satisfying the following condition:
$v_F q < \omega < 
2E_F - v_F q$, in which the undamped plasmon modes exist.  
Since the normalized plasmon mode dispersion in $E_F$ scales with
normalized Fermi wave vector,
our calculated results of Fig. \ref{fig0} apply also for all other
densities.

As shown in Fig.~\ref{fig0}(a), for a small spatial separation ($k_Fd
<1$), the frequency of the acoustic mode $\omega_-$ decreases 
compared with the uncoupled mode and approaches 
the upper boundary of intraband SPE (i.e. $\omega=v_F q$) while the
optical mode $\omega_+$ shifts to higher energy.
Especially, as $d \rightarrow 0$, the $\omega_-$ mode
becomes degenerate with the electron-hole continuum and loses its
identity (i.e. $\omega_- \rightarrow v_F q$). On the other hand,
the $\omega_+$ mode becomes the SLG plasmon with a density
$N= n_1 + n_2 +2\sqrt{n_1 n_2}$ in the long wavelength limit. This is
very interesting because,
when the two layers with  densities $n_1,n_2$
become one combined system ($d \rightarrow 0$), 
the optical plasmon mode ($\omega_+$) should correspond to the SLG 
plasmon of total density $N=n_1+n_2$, which is precisely what happens
in regular 2D double layer system.
As the spatial separation increases the optical
(acoustic) plasmon frequency decreases (increases)
slowly and the mode coupling occurs at the boundary of
interband SPE (see 
Fig.~\ref{fig0}(b)). As $d$ increases further the two modes
become degenerate below the SPE$_{\rm inter}$ (i.e. in the undamped
region), but the degenerate mode is divided into two coupled modes
inside interband SPE 
(see Fig.~\ref{fig0}(c) and (d)).  
For infinite
separation, the two modes become uncoupled, and have the
dispersion of SLG plasmon with the density $n=n_1=n_2$.

\begin{figure}
\bigskip
\epsfxsize=1.02\hsize
\epsffile{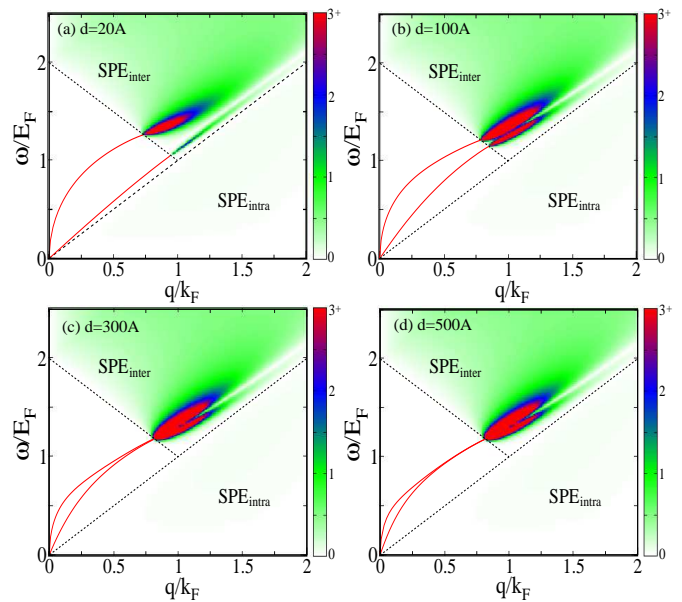}
\caption{\label{fig1}
The density plots of DLG loss function ($-{\rm
  Im}[\epsilon(q,\omega)^{-1}]$) in ($q,\omega$) space for fixed 
densities of $n_1=n_2=10^{12}cm^{-2}$ and several layer
separations $d=20$, 100, 300, 500$\AA$. 
The dashed lines indicate the boundaries of SPEs and
the solid lines represent the
undamped plasmon modes ($\delta$-function peaks in the loss function).
}
\end{figure}

In Fig.~\ref{fig1} we show the calculated  DLG loss function 
(i.e. $-{\rm  Im}[\epsilon(q,\omega)^{-1}]$)  
plotted in $q$-$\omega$ space. The loss function is
related  to the 
dynamical structure factor $S(q,\omega)$ by
$S(q,\omega) \propto -{\rm Im} \left [ 
{\epsilon(q,\omega)}^{-1} \right ]$.
The dynamical structure factor gives a direct 
measure of the spectral strength of the various elementary excitations. 
Thus, our calculated loss function can be measured in
experiments such as inelastic electron  
and Raman scattering spectroscopies.
When both Re$[\epsilon]$ and
Im$[\epsilon]$ become zero (i.e. $\epsilon(q,\omega) = 0$, which
defines the plasmon mode),
the imaginary part of the inverse dielectric function,
Im$[\epsilon(q,\omega)^{-1}]$, is a
$\delta$-function with the strength 
$ W(q)=\pi [|\partial {\rm Re}[\epsilon(q,\omega)] /
\partial \omega|_{\omega=\omega_p(q)}]^{-1}$,
where $\omega_p(q)$ is the plasmon dispersion.
Thus the undamped plasmon shows up as well defined $\delta$-function peak in
the loss function. The damped plasmon, however,
corresponds to a broadened peak in the loss function -- for larger
broadening, the plasmon is overdamped, and there is no peak in the loss
function.

In Fig.~\ref{fig1} we show the $-$Im[$\epsilon(q,\omega)^{-1}$] in
density plots where darkness 
represents the mode spectral strength.
The solid lines  indicate the $\delta$-function peaks 
and correspond to the well-defined undamped plasmon modes. 
The intraband electron-hole SPE continuum
shows up as weak broad (incoherent) structure in Fig.~\ref{fig1} and
carries small spectral weight. 
For a small spatial separation only  the in-phase mode $\omega_+$ carries
any significant spectral strength.
As $d$ increases the out-of-phase mode carries substantial
spectral weight even at long wavelengths ($q \rightarrow 0$).
Both $\omega_{\pm}$
modes in general carry finite spectral weights and should be observable
in experiments \cite{pinczuk,allen,R1,Liu}.
Since both $\omega_{\pm}(q)$ are greater than $\omega = v_F q$ for all
$q$, which is 
the upper boundary of the intraband electron-hole pair continuum, we 
expect these modes not to decay to intraband electron-hole pairs. 
But the modes enter into interband SPE at finite $q_c$, which is given
by the condition, $\omega_{\pm}(q_c) = 2E_F - v_F q_c$.
If $q>q_c$, $\epsilon(q,\omega_{\pm})$ has a finite imaginary part and the
plasmon modes become damped
consequently (Landau damping). The plasmon mode inside
the Landau damping region decays by emitting interband electron-hole pairs
which is now allowed by energy-momentum conservation.
Near the boundary of the interband SPE the mode damping is not
significant.

\begin{figure}
\bigskip
\epsfxsize=1.0\hsize
\hspace{0.0\hsize}
\epsffile{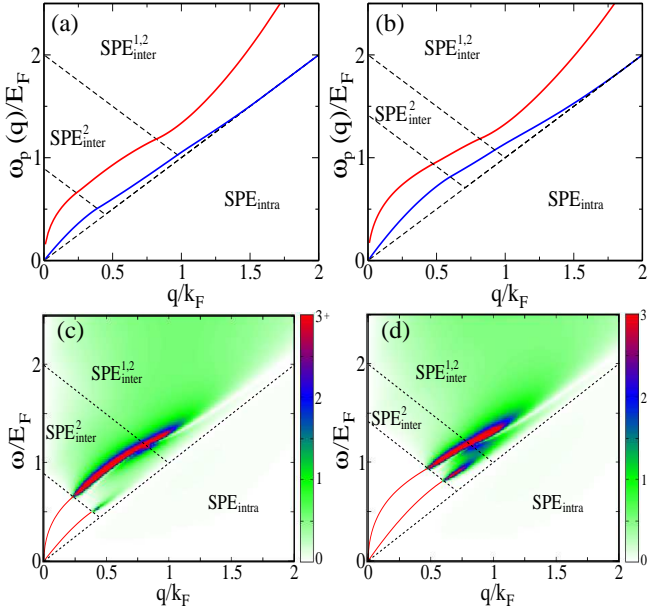}
\caption{\label{fig3}
Calculated plasmon mode dispersions and $-{\rm Im}[\epsilon(q,\omega)^{-1}]$
of DLG for fixed 
layer separation ($d=100 \AA$) and for different layer densities.
Here we use  $n_1=10^{12} cm^{-2}$ and
in (a)(c) $n_2/n_1 = 0.2$ and  in (b)(d) $n_2/n_1 = 0.5$.}
\end{figure}

In Fig.~\ref{fig3} we show the calculated DLG plasmon mode dispersion and 
the corresponding loss function ($-{\rm
  Im}[\epsilon(q,\omega)^{-1}]$) for
unequal layer densities at fixed 
layer separation ($d=100$ \AA).
As the layer densities are different the undamped plasmon region 
becomes smaller in $q$-$\omega$ space. As $q$
increases the modes first enter
SPE$_{\rm inter}^2$ at smaller $q_c$ and  decay by 
producing interband electron-hole pairs in the layer 2. As $q$
increases further the modes enter into SPE$_{\rm inter}^1$
and decay by exciting interband electron-hole pairs in both
layers. As the density
imbalance increases (i.e. $n_2/n_1$ decreases) the acoustic mode
($\omega_-$) approaches  the boundary of intarband SPE and loses
its spectral strength severely. But the 
in-phase mode $\omega_+$ remains a well defined peak inside
SPE$^2_{\rm inter}$.

\begin{figure}
\bigskip
\epsfxsize=1.0\hsize
\hspace{0.0\hsize}
\epsffile{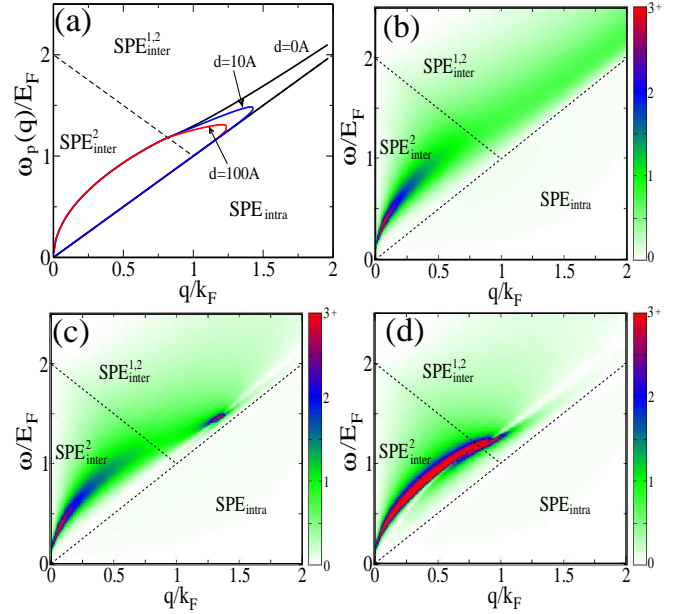}
\caption{\label{fig2}
(a) Calculated plasmon mode dispersions of DLG for
  different layer separations for $n_2/n_1=0$ (i.e. $n_2=0$ and
  $n_1=10^{12}cm^{-2}$), and corresponding loss 
  functions ($-{\rm Im}[\epsilon(q,\omega)^{-1}]$) for (b) $d=0\AA$, (c)
  $d=10\AA$, and (d) $d=100\AA$.
}
\end{figure}

In Fig.~\ref{fig2}(a) we show the DLG plasmon modes 
in the case of $n_2/n_1 =0$, i.e., the layer 2 is undoped
($n_2=0$) and the layer 1 has a 
finite density $n_1=10^{12}cm^{-2}$. Even though there are no free carriers
in layer 2 we have two coupled plasmon modes (i.e we find two zeros in
Re[$\epsilon(q,\omega)$]): one ($\omega_+$) is above SPE$_{\rm
  intra}$ and the other ($\omega_-$) is degenerate with the boundary
of SPE$_{\rm   intra}$ (i.e.  $\omega(q)=v_F q$).
More interestingly, when the layer separation is finite (i.e. $d\neq 0$)
the two plasmons merge at a finite wave vector $q_m$ ($ > k_F$). We
find that the mode
dispersion and $q_m$ are unchanged when $d >100$ \AA. As
$d \rightarrow 0$  
the dispersion of $\omega_+$ becomes exactly that
of SLG plasmon with the same density, and  
the two modes $\omega_{\pm}$ never meet.
In Fig.~\ref{fig2}(b)-(d) we show the loss function of DLG
corresponding to the Fig.~\ref{fig2}(a).
Since all modes are inside of SPE$_{\rm inter}^2$ there is no
undamped Landau region, so the modes
are damped by producing interband electron-hole pairs in layer 2 even
in the long 
wavelength limit. The loss function shows that  $\omega_-$
does not carry any spectral weight, and
all spectral weight is carried by $\omega_+$.
For $d=0$, the $\omega_+$ mode becomes a well defined peak only in the long
wavelength limit and appears as a very broad peak for high
wave vectors.
As the separation increases, $\omega_+$ 
is well defined below $q_m$ even inside
SPE$_{\rm inter}$ and manifests well defined spectral peaks.
Note that there is no significant spectral weight for $q>q_m$.

In conclusion, 
we investigate theoretically the
dynamical response of DLG.
We find interesting novel plasmon features
unique to DLG which do not exist in ordinary 2D double
quantum well structures.

This work is supported by DOE-Sandia and US-ONR. We acknowledge the
hospitality of the KITP at UCSB where this research was initiated.

\end{document}